\documentstyle[12pt]{article}

\begin{document}
\titlepage

\title{Finiteness of multi-loop superstring amplitudes}
\author{G.S. Danilov\thanks{E-mail address:
danilov@thd.pnpi.spb.ru}\\
Petersburg Nuclear Physics Institute\\
Gatchina, St.-Petersburg 188350, Russia}
\date{}
\maketitle

\begin{abstract}
Superstring amplitudes of an arbitrary genus are calculated
through super-Schottky parameters by a summation over the
fermion strings. For a calculation of divergent multi-loop
fermion string amplitudes a supermodular invariant regularization
procedure is used. A cancellation of divergences in the
superstring amplitudes is established. Grassmann variables are
integrated, the superstring amplitudes are obtained to be
explicitly finite and modular invariant.
\end{abstract}

\newpage

\section{Introduction}
During years, a great deal of efforts in superstring theory
\cite{rnshw} has been invested [2---10] in a construction of a
perturbation series for interaction amplitudes. Especially,
difficulties arose in a calculation of partition functions and
of field vacuum correlators for Ramond strings where the desired
values can not be derived by an obvious extension of boson
string results \cite{vec7}. Finally, by a method given in
\cite{dan1,danphr} the partition functions and the superfield
vacuum correlators were calculated \cite{danphr,dan6} in terms
of super-Schottky parameters for all the fermion strings.
Multi-loop superstring amplitudes could be obtained by a
summation of the fermion strings, but every fermion string
amplitude is divergent. Though the divergences are expected
\cite{ver,martpl,berk} to be canceled in the full superstring
amplitude, up to now they kept the desired superstring
amplitudes from being calculated. In this paper a supermodular
invariant calculation \cite{dprpr} of divergent fermion string
amplitudes is proposed. So, superstring amplitudes calculated by
a summation over the fermion string ones are surely invariant
under supermodular group.  We establish a cancellation of
divergences in the above superstring amplitudes. Moreover, on
integration over Grassmann moduli we obtain expressions for the
superstring amplitudes that are explicitly finite and
supermodular invariant. Details of this construction are need
yet to be clarify,  but the paper mainly completes a building of
the superstring perturbation series that, in turn, opens
opportunities for a wide investigation of superstrings.
A topic of an essential interest that can be
advanced in the next future is a summation of the superstring
perturbation series in an infrared energy region of interacted
states. Amplitudes of genus going to infinity dominate in this
case, the discussed infrared asymptotics are expected to be
quite different from each of multi-loop amplitudes taken in the
infrared limit.  Perhaps, they can be
applied to particle interactions below the Plank mass. Another
goal could be a summation of the superstring perturbation series
for massless string states provided that both a number thereof
and their energies tend to infinity. It might be applied to a
creating of the Universe.

As it is usually, a genus-$n$ closed
superstring amplitude $A_{n,m}$ with $m$ legs is given by
\begin{equation}
A_{n,m}=\int{\prod_N}^\prime dq_Nd\overline
q_N\prod_{r=1}^m dt_rd\overline t_r \sum_{L,L'}
Z_{L,L'}^{(n)}(\{q_N,\overline q_N\})
E_{L,L'}^{(n,m)}(\{t_r,\overline t_r\},\{q_N,\overline
q_N\};\{p_j\})
\label{ampl}
\end{equation}
where $\{p_j\}$ are particle momenta, the overline
denotes the complex conjugation and $L$ ($L'$) labels superspin
structures of right (left) superfields defined on the complex
$(1|1)$ supermanifolds \cite{bshw}. Every superspin structure
$L=(l_1 ,l_2)$ presents a superconformal extension of the
$(l_1,l_2)=\bigcup_{s=1}^n(l_{1s},l_{2s})$ ordinary spin one
\cite{swit}. The genus-$n$ theta function characteristics
$(l_1,l_2)$ can be restricted by $l_{is}\in(0,1/2)$. The prime
denotes a product over those $(3n-3|2n-2)$ super-Schottky
parameters $\{q_N\}$ that are chosen as moduli,
$(3|2)$ thereof being fixed common to all the genus-$n$
supermanifolds by a super-M{\"o}bius transformation. Partition
functions $Z_{L,L'}^{(n)}(\{q_N,\overline q_N\})$ are calculated
from equations \cite{dan1,danphr} expressing that the
superstring amplitudes are independent of a choice of two-dim.
metrics and of a gravitino field. The vacuum expectation
$E_{L,L'}^{(n,m)}(\{t_r,\overline t_r\},\{q_N,\overline
q_N\};\{p_j\})$ of the vertex product is integrated over
supermanifods $t_r=(z_r|\theta_r)$ where $z_r$ is a local complex
coordinate and $\theta_r$ is its odd partner. Moduli are
integrated over the fundamental domain \cite{dnph}.
Among other things, this domain depends on
$L$ through terms proportional to Grassmann super-Schottky
parameters since, generally, the supermodular changes of
moduli and of supercoordinates  depend on superspin structure
\cite{dnph}.

To calculate $A_{n,m}$ a regularization procedure is necessary
because every term in (\ref{ampl}) is divergent \cite{belkniz}
due to degeneration of Riemann surfaces. If a cutoff \cite{grpd}
of modular integrals  is used, it is necessary yet to verify
the supermodular invariance of the calculated amplitudes because
the cutoff \cite{grpd} violates the supermodular group. So we
use a supermodular invariant regularization procedure given in
this paper. Also, we regularize the integrals
over $z_j$ that are ill defined too. A construction of
supermodular covariant functions needed for a regularization of
the modular integrals is complicated by a dependence on the
superspin structure of supermodular changes of super-Schottky
parameters \cite{dnph}. As example, the sum over $(L,L')$ in
(\ref{ampl}) calculated with $\{q_N\}$ common to all the
superspin structures is non-covariant under the supermodular
group though each of the $(L,L')$ terms is covariant \cite{dnph}
under the group considered. Hence we perform a singular
transformation  \cite{dan7} to a new parameterization
$P_{split}$ where transition groups are split, the supermodular
group being reduced to the ordinary modular one. While the
superstring in not invariant under the above transformation, it
is useful to integrate over Grassmann variables because in this
case the integration region is independent of the above
Grassmann ones.

The $P_{split}$ parameterization is considered in Sec.2. In
Sec.3 regularized expressions for superstring amplitudes are
given. It is argued that  amplitudes of an emission of a
longitudinal polarized gauge boson vanish in our scheme as it is
required by the gauge symmetry. In Sec.4 the cancellation of
divergences in superstring amplitudes and non-renormalization
theorems \cite{martpl} are verified. On integration over the
$P_{split}$ Grassmann variables the expressions for the
amplitudes are derived that are free from divergences and
supermodular invariant as well.

\section{Regularization of modular integrals}

As it was mentioned above, we perform a singular
$t\rightarrow \hat t=(\hat z|\hat\theta)$ superholomorphic
transformation \cite{dprpr,dan7} to a $P_{split}$
parameterization where transition groups contain no Grassmann
parameters:
\begin{equation}
z=f_L(\hat z)+f_L'(\hat
z)\hat\theta\xi_L(\hat z),\, \theta=\sqrt{f_L'(\hat
z)}\left[\left(1+\frac{1}{2}\xi_L(\hat z)\xi_L'(\hat z)\right)
\hat\theta+\xi_L(\hat z)\right]\,,\quad f_L(\hat
z)=\hat z+y_L(\hat z)\,.
\label{split}
\end{equation}
Here the "prime" symbolizes $\hat z$-derivative, $\xi_L(\hat z)$
is a Grassmann function and $y_L(\hat z)$ is
proportional to Grassmann modular parameters. On the $t$
supermanifold the rounds about $(A_s,B_s)$-cycles are associated
with super-Schottky transformations
$(\Gamma_{a,s}(l_{1s}),\Gamma_{b,s}(l_{2s}))$ where every
$A_s$-cycle is a Schottky circle. In this case
\cite{dan3,danphr} $\Gamma_{a,s}(l_{1s}=0)=I$,
$\Gamma_{a,s}^2(l_{1s}=1/2)=I$, but
$\Gamma_{a,s}(l_{1s}=1/2)\neq I$. So a square root cut on the
z-plane appears for every $l_{1s}\neq0$ with endcut points to be
inside corresponding Schottky circles. For a handle $s$,
the super-Schottky transformation is determined
by a multiplier $k_s$ and two unmoved points
$(u_s|\mu_s)$ and $(v_s|\nu_s)$ where $\mu_s$ and $\nu_s$ are
Grassmann partners of $u_s$ and, respectively, of $v_s$.
In the $P_{split}$ parameterization the same $(A_s,B_s)$
rounds are associated with transformations
$(\hat\Gamma_{a,s}(l_{1s}),\hat\Gamma_{b,s}(l_{2s}))$. Hence
\begin{equation}
\Gamma_{b,s}(l_{2s})(t)=
t\left(\hat\Gamma_{b,s}(l_{2s})(\hat t)\right)\,,\quad
\Gamma_{a,s}(l_{1s})(t)=
t^{(s)}\left(\hat\Gamma_{a,s}(l_{1s})(\hat t)\right)
\label{main}
\end{equation}
where $t^{(s)}(\hat t)$ is obtained by $2\pi$-twist of $t(\hat
t)$ on the complex $\hat z$-plane about the Schottky circle
assigned to a particular handle $s$. In this case
$\hat\Gamma_{a,s}(l_{1s})$ may only give a sign of fermion
fields and $\hat\Gamma_{b,s}(l_{2s})$ is a Schottky
transformation with a multiplier $\hat k_s$ and two unmoved
local points $\hat u_s$ and $\hat v_s$. So both
$\hat\Gamma_{a,s}(l_{1s})$ and $\hat\Gamma_{b,s}(l_{2s})$ do not
contain Grassmann  modular parameters.
Since every super-Schottky
group depends, among other things, on $(2n-2)$
Grassmann moduli, the transition functions in (\ref{split})
necessary depend on $(2n-2)$ Grassmann parameters
$(\lambda_j^{(1)},\lambda_j^{(2)})$ where $j=1\dots n-1$.
The equations similar to (\ref{main}) were already used
\cite{dnph} in a calculation of the acting of supermodular
transformations on supercoordinates and on modular parameters.
Unlike \cite{dnph}, eqs.(\ref{main}) are satisfied only if the
transition functions in (\ref{split}) have poles in a
fundamental region of $\hat z$-plane, singular parts being
proportional to Grassmann parameters. We take
\footnote{An another choice of the poles is discussed in
\cite{dan7}.} them possessing $(n-1)$ poles $\hat z_j$ of an
order 2. For even superspin structures we choose the above poles
among $n$ zeros of the fermion Green function $R_L^f(\hat z,\hat
z_0)$ calculated for zero Grassmann moduli.\footnote{See Sec. 4
of \cite{danphr} where $R_L^f(\hat z,\hat z_0)$ is denoted as
$R_f(z,z')$.} For odd superspin structures the poles can be
chosen by a similar way \cite{dan7}. We take $\hat z_0$ common
to all superspin structures. In this case supermodular changes
of $(\lambda_j^{(1)},\lambda_j^{(2)})$ are independent of the
superspin structure and the supermodular group in the
$P_{split}$ representation is mainly reduced to the ordinary
modular one. The singular parts of (\ref{split}) are determined
by a condition that above modular group is isomorphic to the
supermodular one in the super-Schottky parameterization.  From
this condition, it is follows \cite{dan7} that near every pole
$\hat z_j(\hat z_0;L)$
\begin{eqnarray}
\xi_L(\hat z)
\approx\frac{[1+ \xi_L(\hat z)\xi_L'(\hat z)]}{R_L^f(\hat z,\hat
z_0)} \left[ \frac{\lambda_j^{(2)} }{[R_L^f(\hat z,\hat
z_0)]}\frac{\partial R_L^f(\hat z,\hat z_0)}{\partial_{\hat
z_0}} +\lambda_j^{(1)}\right]
+\frac{\lambda_j^{(1)}\lambda_j^{(2)}\xi_L(\hat z)}
{2[R_L^f(\hat z,\hat z_0)]^2}
\frac{\partial^2\ln[R_L^f(\hat z,\hat z_0)]}{\partial_{\hat
z}\partial_{\hat z_0}} \,, \nonumber\\
f_L(\hat z)\approx
\frac{\lambda_j^{(2)}
\xi_L(\hat z)f_L'(\hat z)}{[R_L^f(\hat z,\hat z_0)]^2}
\frac{\partial
R_L^f(\hat z,\hat z_0)}{\partial_{\hat z_0}}
+\frac{\lambda_j^{(1)}
\xi_L(\hat z)f_L'(\hat z)}{R_L^f(\hat z,\hat z_0)}
\label{ppart}
\end{eqnarray}
where the "prime" symbol denotes $\partial_{\hat z}$. The
calculation \cite{dprpr,dan7} of both $y_L(\hat z)$ and
$\xi_L(\hat z)$ is quite similar to that in Sec. 3 of
\cite{dnph}. The set of
(\ref{split}) and of (\ref{main}) determines both $t$
and $q_N$ in terms of $\hat t$ and  of $\{\hat q_N\}$ up to
$SL_2$ transformations of $t$  where $\{\hat q_N\}=\{\hat
q_{ev}, \lambda_j^{(1)},\lambda_j^{(2)}\}$ and $\{\hat
q_{ev}\}= \{\hat k_s,\hat u_s,\hat v_s\}$. We consider
the $\{r,j\}$ set of the solutions  fixed by
\begin {equation}
t(\hat t;\{\hat q_N\};L;r,j);\quad
q_N(\{\hat q_N\};L;r,j):\qquad\mu_r=\nu_r=0,\,u_r=\hat
u_r,\,v_r=\hat v_r,\,u_j=\hat u_j\,.
\label{cond}
\end{equation}
Every solution
is obtained by a $SL_2$
transformation $M(r,j;r_0,j_0)$ of the $(r=r_0,j=j_0)$ one as
\begin{equation}
t(\hat t;L;r,j)=M(r,j;r_0,j_0) t(\hat
t;L;r_0,j_0)\,,\qquad \{P(r,j)\}=M(r,j;r_0,j_0)\{P(r_0,j_0)\}\,.
\label{trtr}
\end {equation}
where $\{P(r,j)\}=\{(u_s|\mu_s),(v_s|\nu_s)\}$. The $\{k_s\}$
multipliers are the same for all $(r,j)$. The $P_{split}$
partition functions $\hat Z_{L,L'}^{(n)} (\{\hat
q_N,\overline{\hat q_N}\})$ can be derived by a going to the
$P_{split}$ variables in (\ref{ampl}) as
\begin{equation}
\hat Z_{L,L'}^{(n)}(\{\hat q_N,\overline{\hat
q_N}\})= F_L(\{\hat q_N\};r,j)\overline{F_{L'}(\{\hat
q_N\};r,j)}\tilde Z_{L,L'}^{(n)}(\{q_N,\overline q_N\}) |(\hat
u_r-\hat u_j)(\hat v_r-\hat u_j)|^2
\label{zspl}
\end{equation}
where $F_L(\{\hat q_N\};r,j)$ is the Jacobian of the
transformation and $q_N=q_N(\{\hat q_N\};L;r,j)$. Furthermore,
$\tilde Z_{L,L'}^{(n)}(\{q_N,\overline q_N\})$ being multiplied
by the factor behind it, is just the partition function in
(\ref{ampl}), if $(u_r,v_r,u_j,\mu_r,\nu_r)$ are fixed in
(\ref{ampl}) as in (\ref{cond}) to be common to all genus-$n$
supermanifolds (for details, see eq.(132) in \cite{danphr}).
Under the $SL_2$ transformations (\ref{trtr}) this factor is
re-defined by a factor that arises in the Jacobian due to
parameters of these transformations depend on $\{q_N\}$. As the
result, (\ref{zspl}) appears invariant under the transformations
(\ref{trtr}). In the same way the partition functions in
(\ref{ampl}) being multiplied by the product of the moduli
differentials, are invariant under the discussed transformations.
Supermodular invariant function $Y(\{\hat q_N,\overline{\hat
q_N}\};\hat z_0,\overline{\hat z_0})$ used in a regularization
scheme is constructed as
\begin{equation}
Y(\{\hat q_N,\overline{\hat
q_N}\};\hat z_0,\overline{\hat z_0})=\frac{ [Y_1(\{\hat
q_N,\overline{\hat q_N}\};\hat z_0,\overline{\hat
z_0})]^{2^{n-1}(2^n+1)}} {Y_2(\{\hat q_N,\overline{\hat
q_N}\};\hat z_0,\overline{\hat z_0})}
\label{reg}
\end{equation}
with $Y_1(\{\hat q_N,\overline{\hat q_N}\};\hat
z_0,\overline{\hat z_0})\equiv Y_1$ and $Y_2(\{\hat
q_N,\overline{\hat q_N}\};\hat z_0,\overline{\hat z_0})\equiv
Y_2$ defined to be
\begin{equation}
Y_1= \sum_{L\in\{L_{ev}\}}
\hat Z_{L,L}^{(n)}(\{\hat q_N,\overline{\hat
q_N}\})\quad{\rm and}\quad
Y_2=
\prod_{L\in\{L_{ev}\}}\hat
Z_{L,L}^{(n)}(\{\hat q_N,\overline{\hat q_N}\})
\label{regul}
\end{equation}
where $\{L_{ev}\}$ is the set of $2^{n-1}(2^n+1)$ even spin
structures, $\hat Z_{L,L}^{(n)}(\{\hat q_N,\overline{\hat
q_N}\})$ is defined by (\ref{zspl}) and $\{\hat q_N\}$-set is
common to all superspin structures. Since both $Y_1(\{\hat
q_N,\overline{\hat q_N}\})$ and $Y_2(\{\hat q_N,\overline{\hat
q_N}\})$ receive the same factor under modular transformation of
$\hat q_N$-parameters, the right side of (\ref{reg}) is
invariant under supermodular transformations. In addition, it
tends to infinity, if Riemann surfaces are degenerated. Indeed,
if a particular handle, say $s$, become degenerated, the
corresponding Schottky multiplier $k_s$ tends to zero. In this
case both the nominator and the denominator in (\ref{reg}) tend
to infinity \cite{danphr}, but terms associated with $l_{1s}=0$
have an additional factor $|k_s|^{-1}\rightarrow\infty$ in a
comparison with those associated with non-zero $l_{1s}$. So
$Y(\{\hat q_N,\overline{\hat q_N}\};\hat z_0,\overline{\hat
z_0})\rightarrow\infty$. If a even spin structure of a
genus-$n>1$ is degenerated into odd spin structures, the
partition functions tend to zero \cite{dan6} while not
vanishing, if it is degenerated into even spin ones. So again
$Y(\{\hat q_N,\overline{\hat q_N}\};\hat z_0,\overline{\hat
z_0})\rightarrow\infty$. Hence to regularize the desired
integrals we introduce in the integrand (\ref{ampl}) a
multiplier
\begin{equation}
B_{mod}^{(n)}(\{\hat q_N,\overline{\hat q_N}\};\hat z_0,
\overline{\hat z_0};\delta_0)=\{\exp[-\delta_0
Y(\{\hat q_N,\overline{\hat q_N}\};\hat z_0,\overline{\hat
z_0})]\}_{sym}
\label{modreg}
\end{equation}
where $\delta_0>0$ is a parameter and the right side of
(\ref{modreg}) is symmetrized over all the sets of $(n-1)$ zeros
of the fermion Green function $R_L^f(\hat z,\hat z_0)$.
By the above reasons, (\ref{modreg}) vanishes, if Riemann
surfaces become degenerated that provides the finiteness of the
modular integrals in (\ref{ampl}). The right side of
(\ref{modreg}) is invariant under the $SL_2$ transformations
(\ref{trtr}) of the $\{(u_s|\mu_s),(v_s|\nu_s)\}$ set. In
addition, it is invariant under those $L_2$ transformation of
$\{\hat u_s,\hat v_s\}$ accompanied by a corresponding
$L_2$-transformation of $\hat z_0$ and of
$(\lambda_j^{(1)},\lambda_j^{(2)})$, which
reduce three $(\hat u_r,\hat v_r,\hat
u_j)$ values for particular $(r,j)$ to the fixed ones
$\hat u_r=\hat u_r^{(0)}$, $\hat v_r=\hat v_r^{(0)}$ and
$\hat u_j=\hat u_j^{(0)}$ common to all spin
structures. For a given $\hat t_0=(\hat z_0|\hat\theta=0)$ one
can calculate its image $\tilde t=(z_0|\tilde\theta(z_0))$ under
the mapping (\ref{split}). It is evidently that $\tilde t$ is
defined modulo $L_2$-transformations. In the considered case the
transition functions have no poles because zeros $\hat z_j(\hat
z_0;L)$ of $R_L^f(\hat z,\hat z_0)$ are always different from
$\hat z_0$. Simultaneously, so far as $\hat z_j(\hat z_0;L)$ is
changed under fundamental group transformations,
eqs.(\ref{main}) are satisfied only if every this transformation
is accompanied by an appropriate change of the
$(\lambda_j^{(1)},\lambda_j^{(2)})$ parameters that is
calculated from (\ref{ppart}). Because the above change of
$(\lambda_j^{(1)},\lambda_j^{(2)})$ does not depend on the
superspin structure, (\ref{modreg}) is invariant under the
super-Schottky transformations of $\tilde t$.

\section{Superstring amplitudes}
The integrals over $z_j$ in (\ref{ampl}) are ill defined when all
the vertices tend to coincide, or, alternatively, all they are
moved away from each other. In addition, there is no a region
in the $\{p_jp_l\}$ space where all the nodal domain
integrations giving raise to poles and to threshold
singularities of $A_{n,m}$ would be finite together.  As it is
usual \cite{gsw}, each of the above integrals is calculated at
$Re\,E_j^2<0$ where $E_j$ is a center mass energy in the channel
considered. Then it is extended to $Re\,E_j^2>0$ by an
analytical continuation in $E_j^2$. To regularize the
$t_j$ integrals we need functions depending on two more
supermanifold points $t_a=(t_{-1},t_0)$ in addition to
$\{t_j\}$. One receives in hands the above $t_a$ points
multiplying (\ref{ampl}) by the unity arranged to be a square in
the same integrals, every integral $I_{LL'}^{(n)}=1$ being
\begin{eqnarray}
I_{LL'}^{(n)}=\frac{1}{n}\int\frac{dtd\overline
t}{2\pi i} I_{LL'}^{(n)}(t,\overline t)\,,\quad I_{LL'}^{(n)}
(t,\overline t)= D(t)[J_s(t;L)+\overline{J_s(t;L')}]
[2\pi i\omega(L)-2\pi
i\overline{\omega(L')}]_{sr}^{-1}\nonumber\\ \times
\overline{D(t)}[J_r(t;L)+\overline{J_r(t;L')}]\,,\quad
D(t)=\theta\partial_z+\partial_{\theta}\,.
\label{illin}
\end{eqnarray}
Here $J_s(t;L)$ are the genus-$n$ superholomorphic functions
\cite{danphr} having periods, $D(t)$ is the spinor
derivative and $\omega_{sr}(L)$ is a
supermanifold period matrix dependent on the superspin
structure \cite{pst,danphr}. Due to
$\overline{D(t)}J_r(t;L)=0$, both $\overline{J_s(t;L')}$ and
$J_r(t;L)$ could be omitted, but they are remained to provide
the integrand to have no cuts on the supermanifold.
Integrating (\ref{illin}) by parts one obtains that
$I_{LL'}^{(n)}=1$ as it was announced. With (\ref{illin}), we
define a regularized superstring amplitude $A_{n,m}(\{\delta\})$
with $m>3$ as
\begin{eqnarray}
A_{n,m}(\{\delta\})=\int\left(
{\prod_N}^\prime dq_Nd\overline q_N\right)\left(\prod_{r=1}^m
dt_rd\overline t_r\right)\sum_{L,L'}
Z_{L,L'}^{(n)}E_{L,L'}^{(n,m)}
\left(\prod_{a=-1}^0dt_ad\overline t_a
I_{LL'}^{(n)}(t_a,\overline t_a)\right)
\nonumber \\ \times
B_{mod}^{(n)}(\{q_N,\overline{q_N}\};\hat z_0,\overline{\hat
z_0};\delta_0)
\prod_{(jl)}
B_{jl}^{(n)}(\{t_a,\overline t_a\};\{q_N,\overline{q_N}\};
\{\delta_{jl}\};L,L')
\label{reampl}
\end{eqnarray}
where $t_0=(z_0|\theta)$. Both $Z_{L,L'}^{(n)}$ and
$E_{L,L'}^{(n,m)}$ are the same as in (\ref{ampl}),
the arguments being omitted for brevity. The $(jl)$ symbol
labels pairs of the vertices, $\delta_{jl}>0$ are parameters and
$\{\delta\}=(\delta_0,\{\delta_{jl}\})$. Further, $\hat z_0=\hat
z_0(z_0)$ is calculated together with its Grassmann partner
$\theta(z_0)$ from (\ref{split}) taken at $\hat\theta=0$,
$z=z_0$ and $\theta=\theta(z_0)$. At $\{\delta_{jl}>0\}$ every
factor in the $(jl)$ product tends to zero at
$|z_j-z_l|\rightarrow0$ and at $|z_j-z_l|\rightarrow\infty$.
Explicitly they are given in \cite{dprpr}. The
superstring amplitude $A_{n,m}$ is defined as
$A_{n,m}(\{\delta\rightarrow0\})$ calculated in line with the
usual analytical continuation procedure \cite{gsw} for the
integrals over nodal domains giving rise to poles and threshold
singularities of $A_{n,m}$. The $(3|2)$ super-Schottky
parameters are no moduli, say, they are $\mu_{r_0}=\nu_{r_0}=0$,
$u_{r_0}=u_{r_0}^{(0)}$, $v_{r_0}=v_{r_0}^{(0)}$ and
$u_{j_0}=u_{j_0}^{(0)}$ common to all the supermanifolds. So,
$\{(u_s|\mu_s),(v_s|\nu_s)\}=\{P(r_0,j_0)\}$.
Due to the previous Section, $\{\hat q_N\}$ for every superspin
structure $L$ can be calculated as $\hat q_N=\hat
q_N(\{q_N(r,j)\};L;r,j)$ through any
$\{q_N(r,j)\}=(\{k_s\},\{P(r,j)\}$ where $\{P(r,j)\}$ is
obtained by a transformation (\ref{trtr}) of $\{P(r_0,j_0)\}$.
The result is independent of the choice of $(r,j)$.

The integrations being well defined,
(\ref{reampl}) can be rearranged by a suitable
$SL_2$-transformation $\tilde M$ to the integral over all
$(3n|2n)$ super-Schottky parameters and over $(m-3|m-2)$ values
among $\{(z_j|\theta_j)\}$, the rest being fixed as
$\{z_b\}=(z_1=z_1^{(0)},z_2=z_2^{(0)},z_3=z_3^{(0)})$,
$\theta_1=\theta_2=0$ as
\begin{eqnarray}
A_{n,m}(\{\delta\})=\sum_{L,L'}\int\left(
\prod_{N} dq_Nd\overline q_N\right)dt_0d\overline t_0
\tilde Z_{L,L'}^{(n)}(\{q_N,\overline q_N\})
K_{L,L'}^{(n,m)}(\{z_b\},\{q_N,\overline q_N\},\hat
z_0,\overline{\hat z_0},\{p_j\})\nonumber\\
\times B_{mod}^{(n)}(\{\hat q_N,\overline{\hat q_N}\};\hat
z_0,\overline{\hat z_0};\delta_0)
I_{LL'}^{(n)}(t_0,\overline t_0)
\label{reampl1}
\end{eqnarray}
where the $\tilde Z_{L,L'}^{(n)}(\{q_N,\overline q_N\})$
partition function is symmetrical in the super-Schottky
parameters and the factor just behind it is given by
\begin{eqnarray}
K_{L,L'}^{(n,m)}(\{z_b\},\{q_N,\overline q_N\},\hat
z_0,\overline{\hat z_0},\{p_j\})=
\int\left(\prod_{r=4}^m
dz_rd\overline z_r\right)\left(\prod_{r=3}^m
d\theta_rd\overline\theta_r\right)
dt_{-1},d\overline{t_{-1}}E_{L,L'}^{(n,m)}
\nonumber\\ \times
|(z_1^{(0)}-z_3^{(0)})(z_2^{(0)}-z_3^{(0)})|^2
I_{LL'}^{(n)}(t_{-1}\overline{t_{-1}})
\prod_{(jl)}
B_{jl}^{(n)}(\{t_a,\overline t_a\}\{\delta_{jl}\};L,L')
\label{kk}
\end{eqnarray}
where $E_{L,L'}^{(n,m)}$ is the same as in (\ref{reampl}) and
the factor between $E_{L,L'}^{(n,m)}(\{t_j,\overline t_j\})$ and
$I_{LL'}^{(n)}(t_{-1}\overline{t_{-1}})$ is due to the fixing of
the $\{z_b\}$ set. The modular parameters in (\ref{reampl1}) are
integrated over the fundamental domain \cite{dnph} that is
invariant under $SL_2$ transformations. In addition, they are
restricted by both $z_1^{(0)}$, $z_2^{(0)}$ and $z_3^{(0)}$ to
be outside all the Schottky circles. In (\ref{reampl1}) the
$k_s$ multipliers are the same as in (\ref{reampl}) and the
$\{(u_s|\mu_s),(v_s|\nu_s)\}=\{P\}$ set is related with
$\{P(r_0,j_0)\}$ in (\ref{reampl}) by the
$SL_2$-transformation $\tilde M$ as $\{P(r_0,j_0)\}=\tilde M
\{P\}$. Just as in (\ref{reampl}), the $\{\hat q_N\}$ set
is calculated in term of $\{P(r,j)\}$ given through
$\{P\}$ by
\begin{equation}
\{P(r,j)\}=M(r,j;r_0,j_0)\{P(r_0,j_0)\}=M(r,j;r_0,j_0)
\tilde M\{P\}
\label{rel1}
\end{equation}
where $M(r,j;r_0,j_0)$ is defined in (\ref{trtr}).
Parameters of the transition matrix in (\ref{rel1}) depend on
the super-Schottky parameters assigned to the $r$ handle and on
$(\hat u_r,\hat v_r)$ in (\ref{cond}), but (\ref{reampl1})
is independent of $(\hat u_r,\hat v_r)$ due to $L_2$ symmetry
discussed just below eq.(\ref{modreg}).

In (\ref{reampl1}), after a suitable
rewriting of the integrals over the nodal domains the
regularization factors $B_{jl}^{(n)}(\{t_a,\overline
t_a\}\{\delta_{jl}\};L,L')$ can be removed from the integral.
Hence the gauge symmetry inherent to massless modes presents
though in $A_{n,m}(\{\delta\})$ it is violated due to these
factors.

\section{ Finiteness of the superstring amplitudes}

Divergences due to a degeneration of a handle are already known
\cite{vec8,dnph} to be canceled in the superstring amplitudes.
Additional divergences could be when clusters $Cl$ of handles
arise, the sizes being small compared with distances to
vertexes (except may be to a solely dilaton-vacuum transition
vertex). In this case, however, leading divergences in $A_{n,m}$
disappear due to integrations in (\ref{reampl1})
over Grassmann modular parameters associated with the $Cl$
cluster above. Indeed, here a dependence on modular parameters
$\{q_{N_1}\}$ of the $Cl$ cluster is factorized in the partition
functions.
Besides, when all the vertexes are separated from the $Cl$
cluster the integrand (\ref{kk}) ceases to depend on
$\{q_{N_1}\}$ except only on the limiting point $u_0$, which the
$Cl$ cluster is contracted to. This $u_0$ dependence in
(\ref{kk}) is removed by a boost of the vertex co-ordinates and
of the modular parameters of the remainder. If in $u_0$ the
dilaton-vacuum transition vertex is situated, an additional
$\{q_{N_1}\}$ dependence arises in (\ref{kk}) solely as
the $I_{LL'}^{(n)}(t,\bar t)$
factor (\ref{illin}). Owing to the above structure of the
integrand (\ref{reampl1}), two Grassmann parameters associated
with the $Cl$ cluster, say,
$(\mu_r,\nu_r)\in\{q_{N_1}\}$, are removed from the integrand
(\ref{reampl1})  by an $SL_2$ transformation $M_r$ of
$\{P_{N_1}\}=\{(u_s|\mu_s),(v_s|\nu_s)\}\in\{q_{N_1}\}$
as $\{\tilde P_{N_1}\}=M_r\{\tilde P_{N_1}\}$ where
$\{\tilde
P_{N_1}\}\{(\tilde u_s|\tilde\mu_s),(\tilde
v_s|\tilde\nu_s)\}\in\{q_{N_1}\}$. The desired $M_r$ has a form
(\ref{split}) with transition functions $f_r(z)$ and $\xi_r(z)$
instead of $f_L$ and $\xi_L$ where
\begin{equation}
M_r:\qquad
f_r(z)=z+\mu_r\nu_r\frac{(z-\tilde u_r)}{(\tilde u_r-\tilde
v_r)}\,, \quad\xi_r(z)= \frac{\mu_r(z-\tilde v_r)-
\nu_r(z-\tilde u_r)}{\tilde u_r-\tilde v_r}.
\label{mob}
\end{equation}
So, $\tilde\mu_r=\tilde\nu_r=0$, $u_r=\tilde u_r$ and
$\tilde v_r=v_r+\mu_r\nu_r$. By (\ref{mob}) we go to the
integration over $\{\tilde u_s,\tilde\mu_s,\tilde v_s,
\tilde\nu_s\}\in\{q_{N_1}\}$ and over $(\mu_r,\nu_r)$ as well.
The partition function in (\ref{reampl1}) has the form
\begin{equation}
\tilde Z_{L,L'}^{(n)}(\{q_N,\overline q_N\})=
Z_{inv}^{(n)}(\{q_N,\overline
q_N\};L,L')\prod_{s=1}^n(u_s-v_s-\mu_s\nu_s)^{-1}
\label{pfun}
\end{equation}
with $Z_{inv}^{(n)}(\{q_N,\overline q_N\};L,L')$ to be $SL_2$
invariant \cite{vec8,dnph}. So in the new variables the
$(\mu_r,\nu_r)$ dependence in (\ref{pfun}) is canceled by that
in the Jacobian of the transformation
(\ref{mob}). Due to (\ref{rel1}), the $\{\hat q_{N_1}\}$
set associated with the cluster, can be calculated through
the corresponding super-Schottky multipliers and the
corresponding $\{P(r,j)\}$-variables, which, in turn, are
calculated through $\{\tilde P_{N_1}\}$ as
$M(r,j;r_0,j_0)\tilde MM_r\{\tilde P_{N_1}\}$.
Since $(\mu_r=0,\nu_r=0)$ in both $\{P(r,j)\}$ and
$\{\tilde P_{N_1}\}$, the $M(r,j;r_0,j_0)\tilde MM_r$
transformation is a properly M{\"o}bius one, its parameters
depend only on $(u_r,v_r,\tilde u_r,\tilde v_r)$, the
regularization factor becomes independent of $(\mu_r,\nu_r)$.
Corrections to the partition functions and to (\ref{kk}) are
found to be quadratic in a size of the $Cl$ cluster that is
sufficient to provide a finiteness of the $A_{n,m}$ superstring
amplitudes. So, the discussed divergences are cancelled in every
superspin structure unlike those \cite{vec8,dnph} due to a
degeneration of a handle. In the above consideration only the
invariance of (\ref{modreg}) under the $SL_2$ transformations
(\ref{mob}) and (\ref{trtr}) is used and the supermodular
invariance of (\ref{modreg}) only provides the supermodular
invariance of $A_{n,m}$. Otherwise the particular form
(\ref{modreg}) of the regularization factor is unessential.

The 0-, 1-, 2- and 3- point functions are calculated
by a factorization of the amplitudes in suitable
regions of the integration variables. The above finiteness of
the amplitudes means that 0-, 1-, 2- and 3-point functions of
massless superstring modes vanish as it is expected
\cite{martpl}.

If zeros of certain of the $R_L^f(\hat z,\hat z_0)$ fermion
Green functions go closely each to other, the Jacobian in
(\ref{zspl}) go to infinity that may origin in (\ref{modreg})
singular terms proportional to powers of $\delta$. Generally,
it might give rise additional divergences in $A_{n,m}$. We
found that the discussed terms absent in two-loop amplitudes,
but for arbitrary genus this matter remains to be
seen. The above terms would be supermodular covariant in
themselves because the $P_{split}$ modular group is split and,
so, every term in an expansion of an exponent (\ref{modreg}) in
powers of $(\lambda_j^{(1)},\lambda_j^{(2)})$ is modular
invariant. So, may be, these terms do not appear at all, but,
contrary, may be, they are necessary for the unitarity
conditions. In this case the above terms are naturally expected
to give a finite contribution to $A_{n,m}$ because they correct
finite contributions to the unitarity.

Once in (\ref{reampl1}) the integration over the $P_{split}$
Grassmann variables is performed, the regularization factor can
be removed from the integrand whereas the obtained expressions
are explicitly finite and modular invariant. The
$P_{split}$ description is convenient for Grassmann
integrations because in this case the integration region does
not depend on the Grassmann variables. The transition from
$(t,q_N)$ to the $P_{split}$ variables $(\hat t,\hat q_N)$ was
discussed in Section 2. By $L_2$-transformation of $\hat t$ one
can fix $(\hat z_1,\hat z_2,\hat z_3)$ to be the same as the
$\{z_b\}$ set. Being proportional to
$\{\lambda_j^{(1)},\lambda_j^{(2)}\}$, a difference between $t$
and $\hat t$ can not contribute to the integrals over
supermanifold. Indeed, the above integrals surely do not depend
on the above Grassmann parameters, if they are considered as
functions of $q_N$. So, the transition to $P_{split}$ in
(\ref{reampl1}) lies in substitution $q_N$ through $\hat q_N$
and in initiation of the Jacobian. The integration region over
the supermanifold is determined now by the
$\{\hat k_s,\hat u_s,\hat v_s\}$ parameters of the Schottky
circles. The modular domain is determined by the ordinary
modular group under the condition  $\{z_b\}$ to be outside all
the Schottky circles. So only those terms contribute to the
integral, which contain the product of all the $P_{split}$
Grassmann parameters. Among these terms we distinguish terms
calculated at all $(\lambda_j^{(1)},\lambda_j^{(2)})$ in
$B_{mod}^{(n)}(\{q_N,\overline{q_N}\};\hat z_0,\overline{\hat
z_0};\delta_0)$ to be zeros. Of the terms to be distinguished,
we consider that calculated at all
$(\lambda_j^{(1)},\lambda_j^{(2)})$ in the Jacobian to be zeros.
Since the product of all the super-Schottky modular parameters is
the same as the inverse Jacobian multiplied by the product of
the $P_{split}$ Grassmann modular ones, this term is obtained by
the $(\mu_s,\nu_s)$ differentiation when $\{k_s,u_s,v_s\}$ to be
unchanged and
$B_{mod}^{(n)}(\{q_N,\overline{q_N}\};\hat z_0,\overline{\hat
z_0};\delta_0)$ is not differentiated. The term of interest
is divided in a sum of the term obtained by
the $(\mu_s,\nu_s)$ differentiation under fixed
$\{k_s,u_s,v_s+\mu_s\nu_s\}$ and of the remainder that presents
total derivatives in respect to every $v_s$. The above remainder
we refer to the group of terms that are due to the
$\{\lambda_j^{(1)},\lambda_j^{(2)}\}$ dependence of the Jacobian
and of $\{k_s,u_s,v_s\}$ when they are
expressed through the $P_{split}$ variables. All they
are total derivatives in respect to $\{\hat k_s,\hat
u_s,\hat v_s\}$. Indeed, since
these terms did not contain the product of all the
Grassmann super-Schottky modular parameters, they can
contribute to the integral only as surface terms due to the
integration region to depend on the above Grassmann modular ones.
Hence, if in the new variables the above terms begin to depend
on all the Grassmann variables, they necessary contribute to the
integral as total derivatives. So, the integration over
$P_{split}$ Grassmann variables being preformed, every the
$(L,L')$ term in the integrand of (\ref{reampl1}) appears to be
\begin{equation}
B_{mod}^{(n)}
\left[\left(\prod_s\partial_{\mu_s}\partial_{\bar\mu_s}
\partial_{\nu_s}\partial_{\bar\nu_s}\right)Z_{L,L'}^{(n)}
K_{L,L'}^{(n,m)}I_{LL'}^{(n)}\right]_{\{k_s,u_s,v_s
+\mu_s\nu_s\}}
+B_{mod}^{(n)}\sum_{\hat q}
\partial_{\hat q}H_{\hat q}^{(n,m)}+R
\label{integ}
\end{equation}
where for brevity we omit arguments in the values forming the
integrand in (\ref{reampl1}). The derivatives in the first
term are calculated under fixed the $\{k_s,u_s,v_s+\mu_s\nu_s\}$
variables, which next can be replaced by $\{\hat k_s,\hat
u_s,\hat v_s\}$. The second term was discussed just above and
$R$ is formed by terms proportional to powers of $\delta$ due to
the $\{\lambda_j^{(1)},\lambda_j^{(2)}\}$ dependence of the
regularization factor (\ref{modreg}).

When the first term in (\ref{integ}) is calculated,
the product behind $Z_{inv}^{(n)}(\{q_N,\overline
q_N\};L,L')$ in (\ref{pfun}) can be taken at
$\{\mu_s=0,\nu_s=0\}$ since only the above $SL_2$ invariant part
of (\ref{pfun}) is differentiated. Hence, if the above $Cl$
clusters  of handles arise, the function associated with the
first term of (\ref{integ}) ceases to depend on certain
Grassmann super-Schottky parameters, the first term in
(\ref{integ}) vanishes in the cases of interest and, therefore,
its contribution to $A_{n,m}$ is finite. This term is not,
however, modular covariant because, generally, the Jacobian of a
supermodular transformation depends on $\{\mu_s,\nu_s\}$ owing
to the non-split property \cite{dnph} of the supermodular group.
Only the sum of the first term and of the second one in
(\ref{integ}) is supermodular invariant, the third term being
supermodular invariant, as it was noted already.
The integration by parts reduces the second term in
(\ref{integ}) to an integral $\tilde R$  due to differentiation
of $B_{mod}^{(n)}$ and to an integral $S$ over that part
$\it\tilde b$ of the moduli region boundary $\it b$, which does
not mapped oneself under modular transformations and $\it b$
can be obtained by modular changes of $\it\tilde b$.
Up to terms that are
supermodular invariant in themselves (if they exist at all) the
integrand of $S$ can be calculated without
using an explicit form of the second term in (\ref{integ}), only
from a condition that changes of $S$ under supermodular
transformations cancel the corresponding changes of the first
term. In this case supermodular changes of the first term are
calculated in terms of changes of super-Schottky parameters
under supermodular transformations preserving the $\{z_b\}$ set
in (\ref{reampl1}). The above supermodular changes of the
super-Schottky parameters are obtained by the transformation
(\ref{rel1}) of those changes of above parameters, which were
calculated in \cite{dnph} (see Sections 3 and 4 of \cite{dnph}).
The resulted boundary integral is obtained finite due to the
finiteness of the first term contribution to $A_{n,m}$. Hence
the regularization factors can be removed from both the
discussed integrals. For genus-2 amplitudes, we found that any
additions to the boundary integral are absent and $\tilde R$ can
be neglected in the $\delta\rightarrow0$ limit, as well as $R$.
It is quite plausible that the same is true in a general case
too. In this case  the desired $A_{n,m}$ amplitude appears to be
\begin{equation}
A_{n,m}=\sum_{L,L'}\int\left( \prod_{r}
dq_rd\overline q_r\right)
\left[\left(\prod_s\partial_{\mu_s}\partial_{\bar\mu_s}
\partial_{\nu_s}\partial_{\bar\nu_s}\right)Z_{L,L'}^{(n)}
K_{L,L'}^{(n,m)}\right]_{\{k_s,u_s,v_s
+\mu_s\nu_s\}} +S
\label{final}
\end{equation}
where the boundary integral $S$ is calculated as it was
discussed just above. The obtained  $A_{n,m}$ amplitude is
finite and supermodular invariant as well.
Since superstrings are non-invariant under
the $P_{split}$ transformation, (\ref{final}) differs
essentially from the expressions in \cite{ver} where a split
property of the supermanifolds has been assumed. Unlike
\cite{ver}, our amplitudes do not depend on the choice of a
basis of the gravitino zero modes. As it was discussed above,
for genus-$n>2$ amplitudes we can not at present exclude
additional terms in $S$ that are supermodular invariant in
themselves and additional suprmodular invariant terms
due to contribution of $R$ and of $\tilde R$.  If
they present or not, it can be clarify by a direct examination
of the second and third terms in (\ref{integ}) that is in
progress now. It seems, however, that the discussed terms may
appear only, if it is dictated by the unitarity conditions. In
this case they expected to be finite because they correct
finite contributions to the unitarity.

The work is supported by grant No. 96-02-18021 from the Russian
Fundamental Research Foundation.


\newpage
\begin{thebibliography}{99}
\bibitem{rnshw}
P. Ramond, Phys. Rev. D3 (1971) 2415;
A. Neveu  and J. H. Schwarz,   Nucl. Phys. B31 (1971) 86;
F. Gliozzi, D. Olive and J. Scherk,   Phys. Lett. 65B (1976) 282;
A. M. Polyakov,
Phys. Lett. B 103 (1981) 210.
\bibitem{ver}
E. Verlinde and H. Verlinde, Phys. Lett.  B 192 (1987) 95;
E.  Martinec, Nucl.  Phys. B 281 (1987) 157;
J. Atick and A. Sen, Nucl. Phys. B. 296 (1988) 157; J. Atick, J.
Rabin and A.  Sen, Nucl. Phys. B 299 (1988) 279;
G. Moore and A. Morozov, Nucl.  Phys. B 306 (1988) 387;
A. Yu. Morozov, Teor.  Mat.  Fiz. 81 (1989) 24;
O. Lechtenfeld and A. Parkes, Nucl. Phys. B 332 (1990) 39;
S. Mandelstam, Phys. Lett. B 277 (1992) 82.
\bibitem{marteo}
M. Martellini and P. Teofilatto, Phys. Lett. B 211 (1988) 293.
\bibitem{martpl}
E.  Martinec,  Phys.Lett. B 171 (1986) 189.
\bibitem{vec8}
P. Di Vecchia, K. Hornfeck, M. Frau, A. Lerda and S. Sciuto,
Phys. Lett.  B 211 (1988) 301;
B. E. W. Nilsson, A. K. Tollst{\'e}n and A. W{\"a}tterstam,
Phys Lett. B 222 (1989) 399.
\bibitem{pst}
J. L. Petersen, J. R. Sidenius and A. K. Tollst{\'e}n,
Phys Lett. B 213 (1988) 30;
Nucl.  Phys.  B 317 (1989) 109;
\bibitem{dan1}
G. S. Danilov, Phys.  Lett. B 257 (1991) 285;
G. S. Danilov, Sov. J. Nucl.  Phys. 52 (1990) 727
[Jadernaya Fizika 52 (1990) 1143];
G. S. Danilov, Physics of Atomic Nuclei 57 (1994) 2183
[Jadernaya Fizika 57 (1994) 2272];
G. S. Danilov, Physics of Atomic Nuclei 58 (1995) 1984
[Jadernaya Fizika 58 (1995) 2095].
\bibitem{dan3}
G.S. Danilov, JETP Lett. 58 (1993) 796 [Pis'ma JhETF 58 (1993)
790.]
\bibitem{danphr}
G. S. Danilov, Phys. Rev. D 51 (1995) 4359 [Erratum: ibid, 1995,
P. 6201].
\bibitem{dan6}
G. S. Danilov, Physics of Atomic
Nuclei 59 (1996) 1774 [Jadernaya Fizika 59 (1996) 1837].
\bibitem{vec7}
P. Di Vecchia, M. Frau, A. Lerda and S.
Sciuto, Phys. Lett. B 199 (1987) 49.
\bibitem{berk}
N. Berkovits, Nucl. Phys. B 408 (1993) 43.
\bibitem{dprpr}
G. S. Danilov, 1997, PNPI-2181, hep-th/9709083
\bibitem{bshw}
M. A. Baranov and A. S.
Schwarz, Pis'ma ZhETF 42 (1985) 340 [JETP Lett. 49 (1986) 419];
D. Friedan, Proc. Santa Barbara Workshop on Unified
String theories, eds. D. Gross and M. Green ( World Scientific,
Singapore, 1986).
\bibitem{swit}
N. Seiberg and E. Witten, Nucl.Phys. B276 (1986) 272.
\bibitem{dnph}
G. S. Danilov, Nucl. Phys. B 463 (1996) 443.
\bibitem{belkniz}
A. A. Belavin and V. G. Knizhnik, Phys. Lett. B 168 (1986) 201.
\bibitem{grpd}
D. J. Gross and V. Periwal, Phys. Rev. Lett. 60 (1988) 2105;
S. Davis, Class. Quantum Gravity, 7 (1990) 1887.
\bibitem{dan7}
G. S. Danilov, Physics of Atomic
Nuclei 60 (1997) 1358 [Jadernaya Fizika 60 (1997) 1495].
\bibitem{gsw}
M.B. Green, J.H. Schwarz and E. Witten, Superstring Theory,
vols.I and II  ( Cambridge Univesity Press, England,
1987).
\end{thebibliography}
\end{document}